%% file: ms.tex
\documentclass[useAMS,usenatbib]{mn2e}
\usepackage{epsfig,color}
\input{macro2}

\title[Autocorrelations of stellar light and mass]
{Autocorrelations of stellar light and mass in the low-redshift Universe}
\author[C. Li and S. D. M. White]
{Cheng Li$^{1,2}$\thanks{E-mail: leech@mpa-garching.mpg.de}
and Simon D.~M. White$^{1}$\\
$^{1}$Max-Planck-Institute for Astrophysics, Karl-Schwarzschild-Str.
1, D-85741  Garching,  Germany   \\ 
$^{2}$MPA/SHAO  Joint   Center for Astrophysical Cosmology at Shanghai
  Astronomical Observatory, Nandan Road 80, Shanghai 200030, China }

\begin{document}

\date{Accepted ........ Received ........; in original form ........}

\pagerange{\pageref{firstpage}--\pageref{lastpage}} \pubyear{2009}

\maketitle

\label{firstpage}

\begin{abstract}
The final data release of the Sloan Digital Sky Survey (SDSS) provides
reliable photometry and spectroscopy for about half a million galaxies
with  median  redshift 0.09.   Here  we  use  these data  to  estimate
projected  autocorrelation  functions  $w_p(r_p)$  for  the  light  of
galaxies  in the  five SDSS  photometric bands.   Comparison  with the
analogous stellar mass autocorrelation, estimated in a previous paper,
shows that stellar luminosity  is less strongly clustered than stellar
mass in  all bands and on  all scales.  Over the  full nonlinear range
$10h^{-1}$kpc$<r_p<10h^{-1}$Mpc  our   autocorrelation  estimates  are
extremely  well represented  by  power laws.   The  parameters of  the
corresponding  spatial   functions  $\xi(r)  =   (r/r_0)^\gamma$  vary
systematically  from  $r_0=4.5h^{-1}$Mpc  and $\gamma=-1.74$  for  the
bluest band  (the $u$  band) to $r_0=5.8h^{-1}Mpc$  and $\gamma=-1.83$
for  the reddest  one  (the $z$  band).   These may  be compared  with
$r_0=6.1h^{-1}$Mpc and $\gamma=-1.84$ for  the stellar mass. Ratios of
$w_p(r_p)$ between  two given wavebands  are proportional to  the mean
colour of correlated stars at projected distance $r_p$ from a randomly
chosen  star.    The  ratio  of   the  stellar  mass   and  luminosity
autocorrelations  measures  an  analogous mean  stellar  mass-to-light
ratio  ($M_\ast/L$).  All  colours  get redder  and all  mass-to-light
ratios get  larger with  decreasing $r_p$, with  the amplitude  of the
effects decreasing strongly to redder passbands. Even for the $u$-band
the effects are quite modest, with maximum shifts of about 0.1 in $u -
g$ and  about 25\%  in $M_\ast/L_u$.  These  trends provide  a precise
characterisation of  the well-known dependence  of stellar populations
on environment.
\end{abstract}


\section{Introduction}\label{sec:introduction}

Over the past three decades, redshift surveys of nearby galaxies have
established that galaxies of different types are distributed in space
in different ways \citep[e.g.][]{Davis-Geller-76, Dressler-80}. Among
the various galaxy properties that have been considered, colour is
found to be among the most dependent on local environment density
\citep{Kauffmann-04, Blanton-05b}, with luminosity also a
significantly environment-dependent property\citep{Blanton-05b}.
Two-point autocorrelation functions, the traditional quantitative
characterisation of clustering, thus depend both on luminosity
\citep{Davis-88, Hamilton-88, White-Tully-Davis-88,
  Boerner-Mo-Zhou-89, Einasto-91, Park-94, Loveday-95, Benoist-96,
  Guzzo-97, Beisbart-Kerscher-00, Norberg-01, Zehavi-02, Zehavi-05,
  Li-06b, Skibba-06, Wang-07, Swanson-08} and on colour
\citep{Willmer-daCosta-Pellegrini-98, Brown-Webster-Boyle-00,
  Zehavi-02, Zehavi-05, Li-06b, Wang-07, Swanson-08,
  Skibba-Sheth-09}.  Measurements of correlation functions for
different classes of galaxies and with different weighting schemes
provide powerful quantitative constraints on models of galaxy
formation and evolution.

The  Sloan Digital  Sky  Survey \citep[SDSS;][]{York-00}  is the  most
ambitious  optical  imaging  and  spectroscopic survey  to  date.   In
\citet[][hereafter Paper  I]{Li-White-09} we studied  the distribution
of stellar  mass in  the local Universe  using a complete  and uniform
sample of about  half a million galaxies selected  from the final data
release \citep[DR7;][]{Abazajian-09} of the SDSS.  This was quantified
by two statistics: the abundance of galaxies as  a function of stellar
mass $\Phi(M_\ast)$,  which we estimated  over the stellar  mass range
$10^{8}<M_\ast<10^{12}h^{-2}$M$_\odot$, and the projected stellar mass
autocorrelation function $w_p^\ast(r_p)$,  which we estimated over the
projected  separation  range  $10 h^{-1}$kpc$<r_p<30h^{-1}$Mpc.   Both
statistics were  robustly and precisely determined for  the masses and
scales  probed.   We  found  $w^\ast_p(r_p)$  to  be  remarkably  well
described by a power law,  a behaviour which is approximately, but not
perfectly  reproduced  by  existing  galaxy  formation  models.  These
measurements  have been  used in  \citet{Guo-09} to  link  the stellar
masses of galaxies to the dark matter masses of their haloes.

In  this short  paper we  extend  the work  of Paper  I by  estimating
projected {\em luminosity}  autocorrelation functions.  We use exactly
the same  methodology and the  same galaxy sample  as in Paper  I.  We
compute luminosity autocorrelation functions for the five passbands of
SDSS, and compare these with the stellar mass autocorrelation function
obtained in  Paper I. For each  case we provide the  parameters of the
best-fitting power-law. By taking  ratios of these autocorrelations we
investigate  the  scale-dependence  of   the  mean  colours  and  mean
mass-to-light  ratios of clustered  stellar populations.   Finally, we
discuss  briefly how  our  results  relate to  other  measures of  the
distribution of stellar light and mass in the low-redshift Universe.

\section{Stellar luminosity autocorrelation functions}

We  use the  same galaxy  sample as  in Paper  I except  that  we have
dropped those  galaxies with  $r$-band absolute magnitude  outside the
range  $-16<M_{^{0.1}r}<-24$ or with  poorly determined  magnitudes in
any of  the other SDSS photometric  bands. This reduces  the sample by
less  than 1\%  (of  which less  than  0.2\% is  due  to rejection  of
galaxies with poor photometry in bands  other than $r$) so that it now
consists of 482,755 galaxies. We use SDSS Petrosian magnitudes so that
for each  galaxy the  luminosities, colours and  stellar mass  are all
measured within  a single well-defined  aperture (defined to  be twice
the  $r$-band  Petrosian  radius).   Our  methodology  for  estimating
projected  autocorrelation functions and  for constructing  the random
sample necessary for such estimates is also identical to that in Paper
I:  the   stellar  masses  of  the  pair   members,  $M_{\ast,i}$  and
$M_{\ast,j}$  in Eqs.   3-5 of  Paper I,  are simply  replaced  by the
corresponding  luminosities, $L_{\alpha,i}$ and  $L_{\alpha,j}$.  Here
$\alpha$ denotes  the passband  being considered which  is one  of the
five bands of SDSS: $u$, $g$,  $r$, $i$, or $z$.  The luminosities are
$K$-corrected       to       their       values       at       $z=0.1$
\citep[see][]{Blanton-03c,Blanton-Roweis-07},  and  are corrected  for
evolution following \citet{Blanton-03a}.

\begin{figure}
\centerline{\epsfig{figure=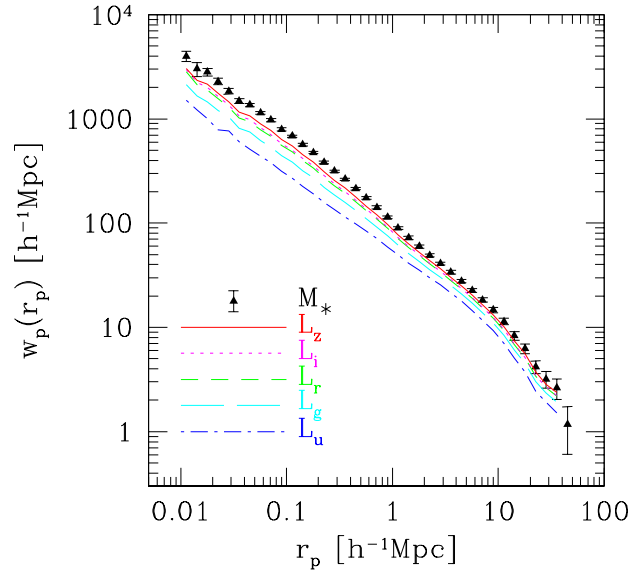,width=0.48\textwidth}}
\caption{The  projected stellar mass  autocorrelation function  in the
  SDSS is plotted as triangles with  error bars and is compared to the
  projected luminosity autocorrelation functions measured for the five
  passbands  of  SDSS  (the   lines).   Errors  on  the  stellar  mass
  autocorrelation function  are estimated  from the scatter  among the
  measurements  from 20  mock galaxy  catalogues constructed  from the
  Millennium Simulation  \citep{Springel-05} using the  same selection
  criteria as the real sample.}
\label{fig:wrp}
\end{figure}

\begin{figure}
\centerline{\epsfig{figure=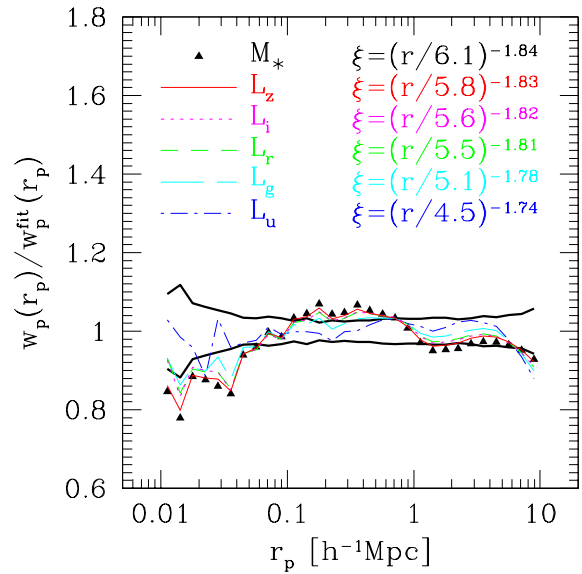,width=0.48\textwidth}}
\caption{Ratio of the projected stellar mass (triangles) or luminosity
  (thin lines) autocorrelation functions in the SDSS to the individual
  best  fit power-laws  over the  range 10  $h^{-1}$kpc $<  r_p  <$ 10
  $h^{-1}$Mpc. The  correlation length  $r_0$ and the  power-law slope
  $\gamma$  of  the  corresponding  three-dimensional  autocorrelation
  function  $\xi(r)=(r/r_0)^\gamma$ are  indicated.   The thick  lines
  above and below unity indicate the $1-\sigma$ scatter of the stellar
  mass  autocorrelation  functions   estimated  from  20  mock  galaxy
  catalogues    constructed    from    the    Millennium    Simulation
  \citep{Springel-05} to  have the same selection effects  as the real
  sample.}
\label{fig:powlaw}
\end{figure}

Our estimates  of the projected  luminosity autocorrelation functions,
$w_p^L(r_p)$,  are plotted  in  Fig.~\ref{fig:wrp} for  the five  SDSS
bands  and are  compared  with the  projected  stellar mass  function,
$w_p^\ast(r_p)$.  Error bars on $w_p^\ast(r_p)$ are estimated from the
scatter between the results found when exactly the same methodology is
applied to our 20 mock SDSS  samples (see Paper I for details).  We do
not attempt to put independent  error bars on the projected luminosity
autocorrelations because  with our technique  the set of  galaxy pairs
used to  estimate each of these  functions is {\it  exactly} the same,
both in  the real data  and in the  mock catalogues. As a  result, the
realisations of sampling  noise, large-scale structure noise (``cosmic
variance'') and background subtraction  noise are identical in all our
autocorrelation estimates.  Only  the colour and stellar mass-to-light
ratio distributions of individual pair members are differently sampled
for  the different  functions. Since  very large  numbers  of galaxies
contribute  to each autocorrelation  estimate and  these distributions
are narrow, they do not contribute significantly to the error budget.

Fig.~\ref{fig:wrp} shows  that luminosity clusters  less strongly than
stellar mass on all scales,  regardless of which waveband we consider.
Furthermore, the amplitude  of clustering increases sytematically with
increasing wavelength on all scales.

Between 10$h^{-1}$kpc and  10$h^{-1}$Mpc the autocorrelation estimates
are all  well described by power  laws. This is shown  more clearly in
Fig.~\ref{fig:powlaw} where we plot the ratio of each to the power law
that best fits the set of  estimates over this $r_p$ range.  Values of
$r_0$   and   $\gamma$   for   the   corresponding   three-dimensional
autocorrelation functions $\xi(r)=(r/r_0)^\gamma$ are indicated in the
figure.   The result  for  the stellar  mass  correlation function  is
plotted as triangles for comparison. Note that these power laws should
be considered  as representations of our  direct $w_p(r_p)$ estimates,
rather  than as  a ``best-fit''  power  law model  for the  underlying
population   correlations.   Estimating   the  latter   would  require
specification of a full clustering  model up to at least fourth order,
as  well as  careful consideration  of the  strong  covariance between
estimates  of $w_p(r_p)$  on different  scales.  The  two  thick lines
indicate  the  $1-\sigma$ scatter  about  the  mean  for stellar  mass
correlation functions estimated from  20 mock SDSS samples constructed
from  the  Millennium  Simulation \citep{Springel-05,Croton-06}.   The
{\it rms} deviations from a power law  are 4, 5, 6, 6, 7 and 7 percent
for the $u$,  $g$, $r$, $i$ and $z$ band  autocorrelations and for the
stellar   mass  autocorrelation,   respectively.    These  are   quite
comparable to the {\it rms} uncertainty of 5 percent which we estimate
for all  these autocorrelations from  our 20 mock catalogues  over the
same $r_p$ range.

There is an  apparent transition in the $w_p(r_p)$  estimates at about
$1 h^{-1}$Mpc.   This is easily understood within  the Halo Occupation
Distribution (HOD) formalism  as a consequence of the  fact that pairs
at larger separations almost all consist of galaxies which are members
of  two different  halos \citep[e.g.][]{Cooray-Sheth-02}.   On smaller
scales  the autocorrelation functions  become steeper  with increasing
wavelength.   This  is  the   regime  where  there  is  a  substantial
contribution from  pairs which  are members of  the same halo  and the
change in shape  may reflect a colour-dependence of  the weights given
to halos of different mass, a colour-dependence of the distribution of
galaxies within a given halo, or (more likely) both.

We have  also estimated  luminosity cross-correlations among  the five
bands. The cross-correlation between two bands is very nearly, but not
exactly,   equal  to   the   geometric  mean   of  the   corresponding
auto-correlations. The  former is  lower than, but  within 2\%  of the
latter  for  all  $r_p$  between  $\sim$50  $h^{-1}$kpc  and  $\sim$20
$h^{-1}Mpc$  and  for   all  band  pairs  that  do   not  involve  the
$u$-band. For cross-correlations involving $u$, the maximum difference
is  just  below  3\%.    Thus,  cross-correlations  do  not  add  much
information   beyond   that   contained  in   auto-correlations   when
characterizing the dependence of stellar populations on environment.

One may  be concerned that systematic effects  in the  SDSS photometry
might significantly bias our correlation  results. We have carried out
two  tests to  investigate   possible  systematics suggested   by  our
referee.   The SDSS  Petrosian magnitudes are  known to  miss a larger
fraction    of  total  galaxy light   in   early-  than  in  late-type
galaxies. Since early-type galaxies  are more strongly clustered, this
might  lead us to underestimate  small-scale clustering. For the first
test we  repeat  our analysis  replacing  Petrosian magnitudes by  the
corresponding  "model  magnitudes" given for each  galaxy  in the SDSS
catalogues. These are estimates of  the total light,  based on fits to
the individual  luminosity   profiles.    The  resulting   correlation
measurements  are  all  higher     than,   but within  3\%    of   the
Petrosian-based values for   all $r_p$ above $\sim$50  $h^{-1}$kpc and
for all bands except $g$.  For the $g$-band, the maximum difference in
the luminosity autocorrelation is still  less than 5\%, and occurs for
$r_p$ between $\sim$100 $h^{-1}$kpc  and $\sim$1 $h^{-1}$Mpc. A second
possible systematic  concerns the well-known    problem that the  SDSS
photometry  tends to   overestimate   the sky   background (and   thus
underestimate the luminosity) for large  objects in crowded fields, in
particular  for brightest cluster galaxies.  Again  this could lead to
systematic underestimates   of small-scale correlations.  We test  the
plausible size  of this effect by  increasing the  luminosities of all
galaxies more  massive  than $10^{11} M_\odot$   by 0.1 mag,  the mean
correction  derived by  \citet{vonderLinden-07}  for  nearby brightest
cluster galaxies (BCGs) larger than 20$^{\prime\prime}$. The resulting
correlation  functions  all remain within  2\%  of the ``uncorrected''
functions on all scales.  Thus, at  least these two systematics appear
likely to affect our results by no more than a few percent.

\section{Colour and stellar mass-to-light ratio distributions}

\begin{figure}
\centerline{\epsfig{figure=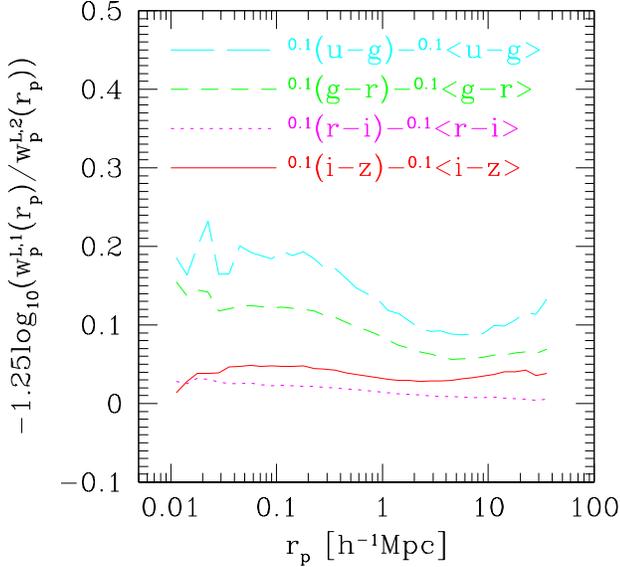,width=0.48\textwidth}}
\caption{Mean colours  of all  correlated stars at  projected distance
  $r_p$   from    a   randomly    chosen   star   relative    to   the
  luminosity-weighted average values of individual galaxies. These can
  be estimated  directly from ratios of  the projected autocorrelation
  functions in Fig.~\ref{fig:wrp}.}
\label{fig:colour}
\end{figure}

If  we take  the  ratio of  our  projected luminosity  autocorrelation
functions in two bands, for example $u$ and $g$, we can write
\begin{eqnarray*}
w^{L_u}_p(r_p)/w^{L_g}_p(r_p))     &     =     &    \frac     {\langle
  L_{u,1}L_{u,2}\rangle_{r_p}   /   \langle   L_u\rangle^2}   {\langle
  L_{g,1}L_{g,2}\rangle_{r_p}  /  \langle   L_g\rangle^2}  \\  &  =  &
\frac{\langle(L_{u,1}/                        L_{g,1})(L_{u,2}/L_{g,2})
  L_{g,1}L_{g,2}\rangle_{r_p}}{\langle     L_{g,1}L_{g,2}\rangle_{r_p}}
\\ & & \div \frac{\langle(L_u/L_g)L_g\rangle^2}{\langle L_g\rangle^2},
\label{eqn:colordef}
\end{eqnarray*} 
where  $\langle ... \rangle_{r_p}$  denotes an  average over  all {\it
  correlated} pairs  of galaxies with projected  separation $r_p$, and
$\langle ... \rangle$ denotes an average over all individual galaxies.
The second equality here shows  that this autocorrelation ratio can be
thought of as the  luminosity-weighted (hence, approximately, per star
weighted) average of the product of the luminosity ratios $L_u/L_g$ of
the pair  members, relative to  the square of  the luminosity-weighted
average  of the  same ratio  for  individual galaxies.   Hence we  can
define  a  characteristic  host  galaxy  colour  for  pairs  of  stars
separated by $r_p$ through, for example,
\begin{equation}
(u        -       g)        -        \langle       u-g\rangle        =
  -1.25\log_{10}(w^{L_u}_p(r_p)/w^{L_g}_p(r_p)).
\end{equation}
This quantity is shown in Fig.~\ref{fig:colour} as a function of $r_p$
for  the four  colour indices  defined  by neighboring  pairs of  SDSS
filters.   The results are  consistent with  host galaxy  colour being
independent of  scale both at  large ($r_p>$ a  few Mpc) and  at small
($r_p<$  a   few  100  kpc)  separations.    The  scale-dependence  at
intermediate separations is  strongest for $^{0.1}(u-g)$ although even
here it amounts to a total variation of less than 0.1 magnitudes.  The
variation in  colour is weaker  for redder colours and  is essentially
absent for the reddest bands.

\begin{figure}
\centerline{\epsfig{figure=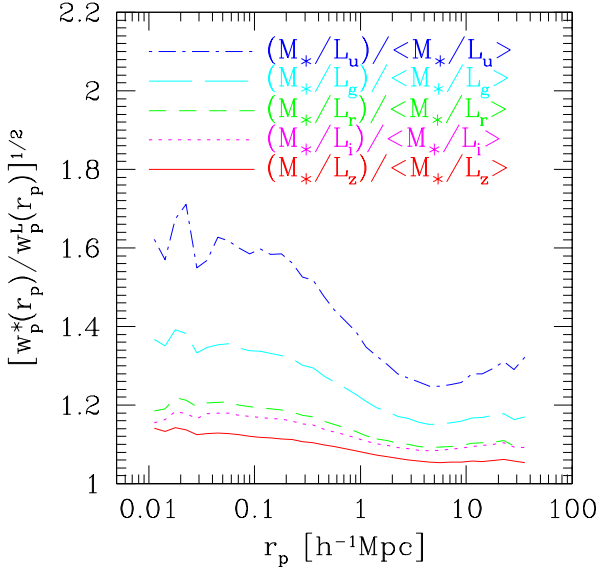,width=0.48\textwidth}}
\caption{Mean  mass-to-light  ratios   for  all  correlated  stars  at
  distance  $r_p$  from  a   randomly  chosen  star  relative  to  the
  luminosity-weighted  average values  for individual  galaxies. These
  can be estimated directly from  ratios of the stellar mass and light
  autocorrelations in Fig.~\ref{fig:wrp}.}

\label{fig:mtol}
\end{figure}

Replacing the luminosity autocorrelation  function in the numerator of
Eq.~\ref{eqn:colordef} by  the projected stellar  mass autocorrelation
function  produces  an  estimate   of  (the  square  of)  the  typical
mass-to-light ratio  of the host  galaxies of (correlated)  star pairs
separated by $r_p$, relative  to the luminosity-weighted average value
for  individual galaxies.   We plot  this quantity  for the  five SDSS
photometric  bands in  Fig.~\ref{fig:mtol}.   The scale-dependence  of
$M_\ast/L$ is  similar to that of  the colours: a  transition to lower
values occurs between  a few 100 kpc and a few  Mpc, but $M_\ast/L$ is
independent of scale on both smaller and larger scales.  Again effects
are strongest in $u$ and get weaker with increasing wavelength, but in
this case a small but significant  trend is seen even in the $z$-band.
$M_\ast/L_u$ increases by about 25\% from large to small scales.

The  trends  shown  in  this  section are  all  easily  understood  as
reflecting  the fact  that  pairs of  small  separation are  typically
members  of the  same halo,  whereas  pairs at  large separation  must
belong to different halos.  This results in a different weighting with
halo mass in the two regimes.  In particular, close (`one halo') pairs
are  more strongly  weighted towards  massive halos  than  are distant
(`two  halo') pairs.   In the  former  case at  least one  of the  two
galaxies must be  a satellite object, whereas in  the latter case both
galaxies are often the central  galaxy of their own halo. These trends
result  in  a greater  contribution  from early-type  ``red-and-dead''
galaxies to the small-scale correlations.

\section{Discussion}

Just as  was the case  for the projected stellar  mass autocorrelation
function presented in Paper  I, projected autocorrelation functions of
luminosity in each  of the five SDSS bands  are robustly and precisely
determined  by  our  methods  for $r_p<30h^{-1}$Mpc.   Over  the  full
nonlinear range  10$h^{-1}$kpc$<r_p < 10h^{-1}$Mpc,  our estimates are
all  extremely  well  represented  by  power  laws,  corresponding  to
three-dimensional auto-correlation functions with parameters $r_0$ and
$\gamma$ which vary slightly but systematically from band to band.  As
pointed out in Paper I, this near power-law behaviour of the estimates
must  be  seen  as   a  coincidence  because  different  physical  and
statistical   processes  control   correlations  in   the  small-scale
(`one-halo')  and large-scale (`two-halo')  regimes.  In  our standard
structure formation model there is no reason why these should conspire
to produce a single power law.  If the ``one-galaxy'' term in the mass
and luminosity autocorrelations were included, there would be a strong
upward  break  at  $r_p\sim  10h^{-1}$kpc,  the  scale  of  individual
galaxies.

Our luminosity autocorrelation estimates  quantify the extent to which
the  spatial  distribution of  optical  light  in  the local  Universe
depends on scale and wavelength. The relative bias between the longest
and shortest  wavebands varies  from a factor  of 1.5 on  small scales
($<$  100kpc) to a  factor of  1.2 on  large scales  ($>$ a  few Mpc).
Luminosity in all bands and  on all scales clusters less strongly than
stellar  mass,  although for  $z$-band  luminosity  the difference  is
small.  This is expected because $z$-band light is known to be closely
related to stellar mass indicators \citep[e.g.][]{Kauffmann-03a}.

All the  autocorrelation functions considered in this  paper have very
similar shape on  large scales ($>5$ Mpc). This  is expected in models
where  the properties  of galaxies  depend on  the  detailed formation
history within their own dark matter halos, but are independent of the
formation histories of  distant halos. On smaller scales  the shape of
the various  autocorrelation functions differ,  reflecting differences
in the  relative strength of  the one-halo and  two-halo contributions
which arise from a  combination of effects.  One-halo correlations are
more  heavily  weighted  towards   massive  halos  than  are  two-halo
correlations, and red and high $M_\ast/L$ galaxies are typically found
in more massive halos than  blue galaxies of similar mass. Small-scale
correlations  are  also  sensitive   to  how  satellite  galaxies  are
distributed    with   radius    within   their    dark    halos   (see
\citealt{Weinmann-06}   and   \citealt{vonderLinden-09}   for   recent
observational studies demonstrating that these radial distributions do
indeed  depend   on  galaxy  colour  and   $M_\ast/L$).   The  precise
quantitative  results we  obtain  in this  paper,  when combined  with
accurate luminosity and stellar  mass functions, provide a compact way
to  constrain  Halo  Occupation   Distribution  models  which  try  to
represent all these relations in detail \citep[e.g.][]{Watson-09}.

The  mass/luminosity auto-correlation  functions  considered here  are
quite closely  related to the marked  correlation functions considered
by others  \citep[MCF; e.g.][]{ Beisbart-Kerscher-00, Faltenbacher-02,
  Sheth-05,   Skibba-06,   Skibba-Sheth-09,   Skibba-09a}.   The   two
statistics  weight data-data  (DD) pairs  in  a similar  way, but  are
usually estimated and  presented differently. In the case  of the MCF,
one doesn't  normally measure the weighted  correlation function (WCF)
in  the way  we do,  but  rather estimates  the ratio  (1+WCF)/(1+UCF)
simply by comparing the weighted DD count with the unweighted one (UCF
here represents the unweighted correlation function). The advantage is
that one  doesn't need to construct  random samples in  order to count
random-random  (RR) and  data-random (DR)  pairs.  A  disadvantage for
projected  correlations of  the  kind  we have  studied,  is that  the
resulting estimate  depends explicitly  not only on  the 3-dimensional
clustering process,  but also on  sample selection procedures  and the
way  the   redshift  separation   of  pairs  is   limited.   Recently,
\cite{Skibba-09a}  have   partially  addressed  this   last  issue  by
estimating weighted  and unweighted projected  CFs, $W_p(r_p,\pi)$ and
$w_p(r_p,\pi)$, integrating them over a fixed range in $\pi$, and then
constructing  the   equivalent  of  a   standard  MCF  as   the  ratio
$[1+W_p(r_p)/r_p]/[1+w_p(r_p)/r_p]$.    As   with   more   traditional
estimates, this  `MCF' goes to unity  on large (linear)  scales and so
tends to obscure interesting information there about the relative bias
of  light, mass or  other `marks',  e.g. the  variation in  colour and
$M_\ast/L$    among     bands    shown    in    Figs.~\ref{fig:colour}
and~\ref{fig:mtol}.

\section*{Acknowledgments}
The authors thank an anonymous referee for helpful comments. CL is
supported by the Joint Postdoctoral Programme in Astrophysical
Cosmology of Max Planck Institute for Astrophysics and Shanghai
Astronomical Observatory, by NSFC (10533030, 10633020), by 973 Program
(No.2007CB815402) and by the Knowledge Innovation Program of CAS
(No.KJCX2-YW-T05).

Funding for  the SDSS and SDSS-II  has been provided by  the Alfred P.
Sloan Foundation, the Participating Institutions, the National Science
Foundation, the  U.S.  Department of Energy,  the National Aeronautics
and Space Administration, the  Japanese Monbukagakusho, the Max Planck
Society,  and the Higher  Education Funding  Council for  England. The
SDSS Web Site is http://www.sdss.org/.

The SDSS is  managed by the Astrophysical Research  Consortium for the
Participating  Institutions. The  Participating  Institutions are  the
American Museum  of Natural History,  Astrophysical Institute Potsdam,
University  of Basel,  University of  Cambridge, Case  Western Reserve
University,  University of Chicago,  Drexel University,  Fermilab, the
Institute  for Advanced  Study, the  Japan Participation  Group, Johns
Hopkins University, the Joint  Institute for Nuclear Astrophysics, the
Kavli Institute  for Particle  Astrophysics and Cosmology,  the Korean
Scientist Group, the Chinese  Academy of Sciences (LAMOST), Los Alamos
National  Laboratory, the  Max-Planck-Institute for  Astronomy (MPIA),
the  Max-Planck-Institute  for Astrophysics  (MPA),  New Mexico  State
University,   Ohio  State   University,   University  of   Pittsburgh,
University  of  Portsmouth, Princeton  University,  the United  States
Naval Observatory, and the University of Washington.

\bibliography{ref}

\bsp
\label{lastpage}

\end{document}

%% file: macro2.tex
\def\reff@jnl#1{{\rm#1\/}}

\def\aj{\reff@jnl{AJ}}                  
\def\araa{\reff@jnl{ARA\&A}}            
\def\apj{\reff@jnl{ApJ}}                
\def\apjl{\reff@jnl{ApJ}}               
\def\apjs{\reff@jnl{ApJS}}              
\def\apss{\reff@jnl{Ap\&SS}}            
\def\aap{\reff@jnl{A\&A}}               
\def\aapr{\reff@jnl{A\&A~Rev.}}         
\def\aaps{\reff@jnl{A\&AS}}             
\def\baas{\reff@jnl{BAAS}}              
\def\jrasc{\reff@jnl{JRASC}}            
\def\memras{\reff@jnl{MmRAS}}           
\def\mnras{\reff@jnl{MNRAS}}            
\def\physrep{\reff@jnl{Phys.Rep.}}
\def\pra{\reff@jnl{Phys.Rev.A}}         
\def\prb{\reff@jnl{Phys.Rev.B}}         
\def\prc{\reff@jnl{Phys.Rev.C}}         
\def\prd{\reff@jnl{Phys.Rev.D}}         
\def\prl{\reff@jnl{Phys.Rev.Lett}}      
\def\pasp{\reff@jnl{PASP}}              
\def\pasj{\reff@jnl{PASJ}}              
\def\skytel{\reff@jnl{S\&T}}            
\def\solphys{\reff@jnl{Solar~Phys.}}    
\def\sovast{\reff@jnl{Soviet~Ast.}}     
\def\ssr{\reff@jnl{Space~Sci.Rev.}}     
\def\nat{\reff@jnl{Nature}}             